# Multi-interface engineering to realize all-solution processed highly efficient Kesterite solar cells

Licheng Lou[1,2]†, Kang Yin[1,2]†, Jinlin Wang[1,2]†, Yuan Li[1], Xiao Xu[1], Bowen Zhang[1,2], Menghan Jiao[1,2], Shudan Chen[1,2], Tan Guo[1,3], Jiangjian Shi[1]*, Huijue Wu[1], Yanhong Luo[1,2,4], Dongmei Li[1,2,4], Qingbo Meng[1,2,4,5]*

[1]Beijing National Laboratory for Condensed Matter Physics, Renewable Energy Laboratory, Institute of Physics, Chinese Academy of Sciences (CAS), Beijing, P. R. China.

[2]School of Physical Sciences, University of Chinese Academy of Sciences, Beijing, P. R. China.

[3]College of Materials Sciences and Opto-Electronic Technology, University of Chinese Academy of Sciences, Beijing, P. R. China.

[4]Songshan Lake Materials Laboratory, Dongguan, P. R. China.

[5]Center of Materials Science and Optoelectronics Engineering, University of Chinese Academy of Sciences, Beijing, P. R. China.

†These authors contributed equally to this work.

*Corresponding author. Email: qbmeng@iphy.ac.cn, shijj@iphy.ac.cn

**Abstract:**

With the rapid development of Kesterite $Cu_2ZnSn(S, Se)_4$ solar cells in the past few years, how to achieve higher cost-performance ratio has become an important topic in the future development and industrialization of this technology. Herein, we demonstrate an all-solution route for the cell fabrication, in particular targeting at the solution processed window layer comprised of ZnO nanoparticles/Ag nanowires. A multi-interface engineering strategy assisted by organic polymers and molecules is explored to synergistically improve the film deposition, passivate the surface defects and facilitate the charge transfer. These efforts help us achieve high-performance and robust Kesterite solar cells at extremely low time and energy costs, with efficiency records of 14.37% and 13.12% being realized in rigid and flexible Kesterite solar cells, respectively. Our strategy here is also promising to be transplanted into other solar cells with similar geometric and energy band structures, helping reduce production costs and shorten the production cycle (i.e. increasing production capacity) of these photovoltaic industries.

Kesterite $Cu_2ZnSn(S, Se)_4$ (CZTSSe) thin-film solar cells have garnered widespread attention due to their advantages such as abundant elemental reserves, environmental friendliness, and high stability.[1,2] The device conversion efficiency has reached the commercial threshold of 15%,[2] making it a rising star in the photovoltaic field. However, in the mainstream fabrication process of CZTSSe devices, the transparent conductive window layer still relies on ZnO/ITO layers prepared by vacuum sputtering. This not only results in elevated material costs, but also requires a substantial investment in equipment.[1,3,4] To further improve the cost performance of CZTSSe solar cells, it is the best choice to realize an all-solution fabrication process.[5] To this end, people have developed a variety of transparent conductive window layer materials suitable for solution preparation, such as conductive polymers[6], nanometal grids,[7] metal oxide nanocrystals[8] or carbon materials[9]. Among them, the composite film window layer composed of silver nanowires (AgNWs) and ZnO nanoparticles (ZnO-nps), with its excellent transparent conductive performance and convenient film coating process, has been widely used in the field of optoelectronic devices.[10-13] CZTSSe cells based on this composite film window layer have achieved an efficiency of 7.51%, showing great development potential.[14-17] However, this efficiency is still far behind the record 14.9% efficiency achieved by devices with vacuum-processed window layers.[2] The main reason is that there are serious defects and energy level mismatching problems within the composite window layer of AgNWs and ZnO-nps, such as: (1) AgNWs are easily corroded, forming a high density of defect states, increasing the recombination of photogenerated carriers;[18] (2) Organic groups remaining on the surface of ZnO-nps form a large number of electron trapping states, inhibiting carrier extraction;[19,20] (3) The large conduction band offset between ZnO-nps and AgNWs forms an electron transport barrier, hindering carrier collection.[21] Therefore, a lot of research work has been carried out, but single interface regulation strategies often have little effect.[13,22-25] How to synergistically solve the multiple interface defects and energy level mismatching problems of this composite window layer remains huge challenges and is key to achieving an efficiency breakthrough in all-solution-processed CZTSSe cells.

This work systematically studies the impact mechanisms of each interface of the ZnO-nps/AgNWs composite window layer on the performance of CZTSSe cells and proposes a multi-interface synergistic control strategy: (1) Using polyacrylic acid (PAA) as a surface binder to construct a conformal ZnO-nps layer, avoiding the corrosion reaction between silver wires and CdS; (2) Introducing aminomethyl phosphonic acid (AMPA) molecules on the surface of ZnO-nps, passivating the surface hydroxyl defects; (3) Simultaneously, the permanent electric dipole of the AMPA molecules forms an interfacial electric field, further eliminating the carrier transfer barrier at the ZnO-nps/AgNWs interface. Ultimately, based on the above synergistic control strategy, the all-solution-processed CZTSSe cell achieved an important breakthrough in the device efficiency from 7.51% to 14.37%, comparable to the current champion devices based on traditional processes. We believe this work will help accelerate the marketization process of CZTSSe thin-film solar cells. At the same time, this synergistic control strategy also provides an important reference for the solution preparation of window layers for other optoelectronic devices.

## Results

### Device performance of all-solution processed solar cells

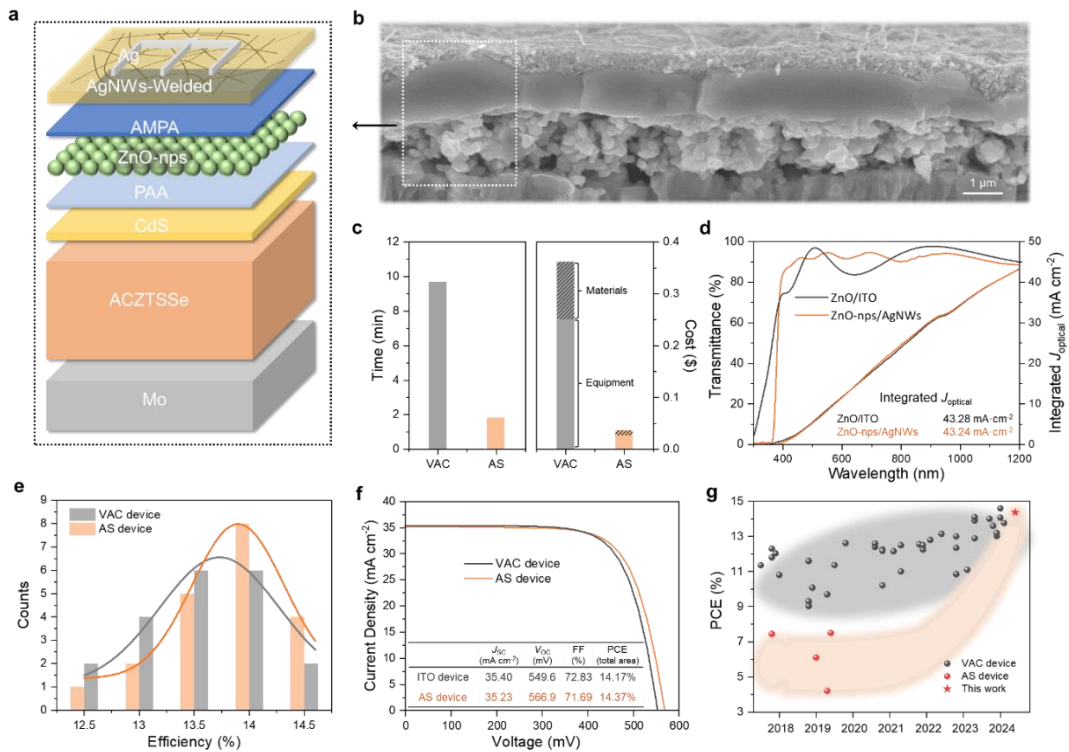

**Figure 1. Performance of the all-solution processed solar cell.** (a) The schematic structure and (b) cross-sectional SEM image of the all-solution fabricated Kesterite solar cell device (AS device). (c) Preparation time and cost for the window layer in a cell fabricated by sputtering or by AS process. The detailed data is shown in Table S1-S2. (d) Transmittance spectrum of sputtered ZnO/ITO and solution fabricated ZnO-nps/AgNWs films, and the corresponding integrated current density ($J_{optical}$) only considering the optical loss. (e) PCE statistical distribution of these solar cells. (f) *I-V* curves of champion solar cells with the window layer fabricated by conventional sputtering method or by our all-solution method. (g) Efficiency evolutions of different types of Kesterite solar cells from 2017. The detailed data is shown in Table S3-4.

The device structure of the Kesterite solar cell in our study is shown in Figures 1a-b, in which all the functional layers were fabricated by solution methods. Specifically, the ACZTSSe and CdS layer were fabricated through the precursor solution coating and chemical bath deposition processes, respectively. The conventional sputtering deposited ZnO/ITO window layer was replaced by ZnO nanoparticles (ZnO-nps) and Ag nanowires (AgNWs) layers deposited by solution coating methods (Figure S1). As in Figure 1b, the ZnO-nps and Ag nanowires with compact morphology and interface contact can be clearly seen in the cross-sectional scanning electron microscope (SEM) image. Finally, screen-printed Ag paste was used as the top grid electrode. We found that using the all-solution preparation route, the time cost for the window layer fabrication can be reduced by about five times and moreover that the energy and material consumption costs can be reduced by nearly 10 times (Figure 1c). In experiment, we have optimized the nano size of ZnO-nps and AgNWs and in the final device 40 nm ZnO-nps and 35 nm AgNWs with a length to diameter ratio of about 700 were used, respectively (Figure S2-S3). Through these material and film deposition process optimization, the composite window layer has exhibited excellent transparent and conductive properties. In a wide wavelength region from visible (400 nm) to infrared (1200 nm), the transmittance of the film was comparable to that of the sputtered ITO film with sheet resistance being controlled <25 Ω/sq (Figure 1d).

In order to overcome the interfacial issues of these solution-processed functional layers, interfaces in the cell were further studied and regulated. Specifically, a thin PAA layer was introduced at the CdS/ZnO-nps interface to gain a better heterogeneous coverage; AMPA molecules were utilized at the ZnO-nps/AgNWs interface to passivate the of ZnO-nps surface defects and moreover facilitate the charge transfer. With these efforts, we have realized higher photoelectric conversion performance in the all-solution processed solar cells compared to that realized by the conventional technology (Figure 1e). In particular, our cell achieved a certified total-area efficiency of 14.37% with short-circuit current density ($J_{SC}$) of 35.23 mA·cm$^{-2}$, open-circuit voltage ($V_{OC}$) of 566.9 mV and fill factor (FF) of 0.7169 (Figure 1f, the corresponding EQE curves is shown in Figure S4), which is almost double that of the previously reported all-solution-processed devices (7.51%, Figure 1g).[14-16] Compared to the sputtered device, our cell here exhibited significantly enhanced $V_{OC}$, which on one hand benefited from the reduced sputtering damage[26] and on the other hand was a result from the reduced interface charge recombination through multi-interface engineering. In the following sections, we will sequentially illustrate the challenges related to fabricating the all-solution processed Kesterite solar cells and how we solved them through the interfacial engineering.

**Engineering of the CdS/ZnO-nps interface**

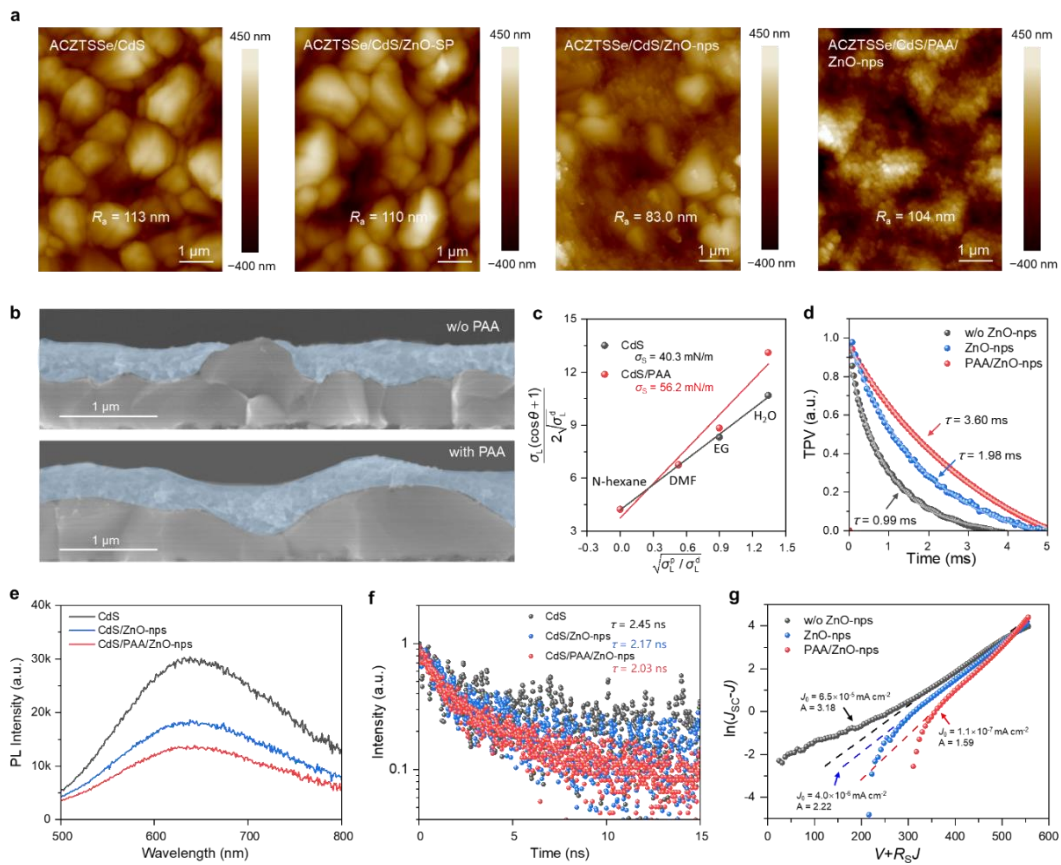

**Figure 2. Engineering of the CdS/ZnO-nps interface.** (a) AFM image of ACZTSSe/CdS, ACZTSSe/CdS/ZnO (sputtered), ACZTSSe/CdS/ZnO-nps, and ACZTSSe/CdS/PAA/ZnO-nps

films. (b) Cross-sectional SEM images of CdS/ZnO-nps interface with or without PAA regulation. (c) Surface energy analysis for CdS and CdS/PAA films. (d) Transient photovoltage of the cell without ZnO-nps, with ZnO-nps and with PAA/ZnO-nps layers. (e) Steady-state and (f) transient photoluminescence of CdS, CdS/ZnO-nps and CdS/PAA/ZnO-nps films excited at 445nm. (g) Idea factor ($A$) and reverse saturation current ($J_0$) of the cells derived from $I$-$V$ curves in Figure S7.[27]

The first challenge is to realize a conformal deposition of ZnO-nps onto the CZTSSe/CdS surface since it usually has large roughness. As shown in Figure 2a-b, when the ZnO-nps directly coated onto the film surface, most nanoparticles were accumulated in the low-lying area of the surface and those towering surfaces can barely be covered, resulting in significant change in surface roughness of the film. This could arise because the ZnO-nps aqueous solution has relatively large polarity and surface tension and thus in its drying process exhibited significant shrink on the CdS surface with low surface energy. This direct exposure of the CZTSSe/CdS surface would not only introduce severe charge recombination but also corrode the AgNWs through Ag-S reactions (Figure S5) [18,28-30].

To address this issue, we introduced a PAA interfacial layer. The PAA is a kind of polymer binder, having rich carboxyl groups which usually have strong interactions with metal oxide nanoparticles.[31,32] Liquid contacting angle measurements found that the PAA obviously enhanced the surface tension,[33] especially the polarity component, of the CdS layer, which thus improved the spread ability of ZnO-nps aqueous solution (Figure 2c, S6, Table S5, Note S1). These effects finally helped the undulating CZTSSe/CdS surface uniformly be covered with ZnO-nps, as shown in Figure 2a-b.

This improvement resulted in an obvious reduction in the interfacial charge recombination, as reflected by the photovoltage decay (Figure 2d).[34,35] In addition, the PL spectra showed that the improvement in the interfacial coverage and contacting also enhanced the carrier transfer at the CdS/ZnO-nps interface. Specifically, CdS in the PAA involved sample (CdS/PAA/ZnO-nps) exhibited a much lower steady-state PL intensity and shorter PL decay lifetime than that of the CdS/ZnO-nps sample (Figure 2e, 2f).[36] Moreover, current density – voltage ($J$-$V$) measurements indicated that the heterojunction property of CZTSSe/CdS/ZnO has also been improved. As in Figure 2g, if CdS was directly contacted with AgNWs window layer, an ideality factor (*A*) as large as 3.18 was observed. Although the direct deposition of ZnO-nps layer decreased this factor to 2.22, it still exceeded the general range of ideality factors for a single heterojunction, i.e. $1 \leq A \leq 2$.[27] Comparatively, when the PAA was used, the ideality factor was reduced to 1.59 with a much lower reverse saturation current density of $1.1 \times 10^{-7}$ mA·cm$^{-2}$, indicating an effective heterojunction was constructed. With these benefits to the CdS/ZnO-nps interface, the cell efficiency was improved from ~10% to ~12.8% (Figure S8).

**Field-effect passivation of ZnO-nps surface defect**

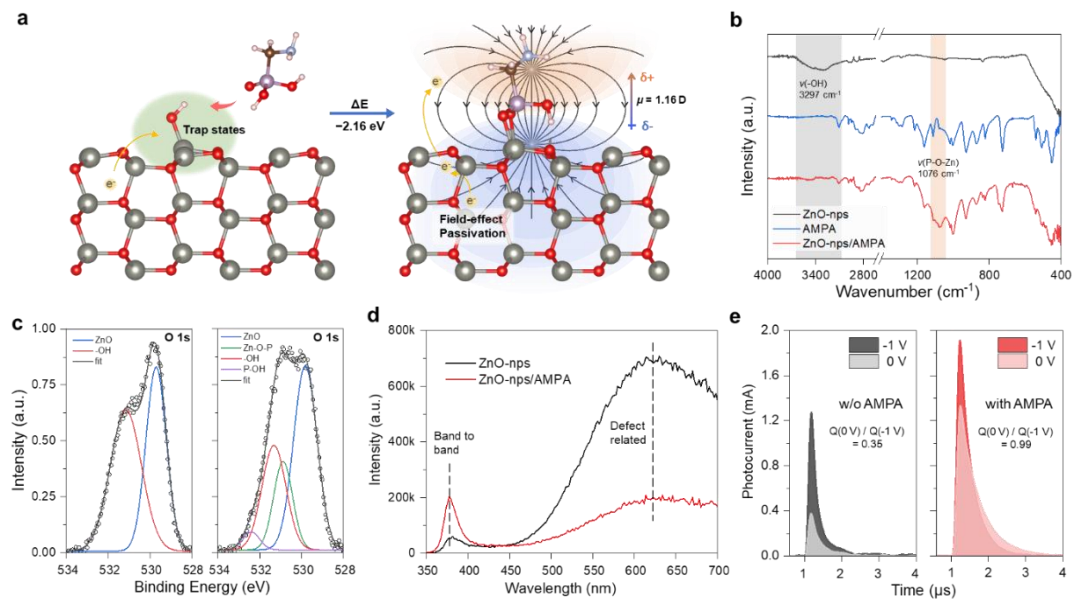

**Figure 3. Field-effect passivation of ZnO-nps surface defect.** (a) DFT simulated reaction between AMPA molecule and -OH on ZnO-nps surface. The black arrow schematically shows the electric field direction of the formed dipole on the ZnO-nps surface after the reaction. (b) FTIR spectra of ZnO-nps, AMPA, and ZnO-nps/AMPA powders collected from films. (c) XPS of O 1s for ZnO-nps and ZnO-nps/AMPA films. (d) Steady state PL spectra of ZnO-nps and ZnO-nps/AMPA films. (e) Bias voltage-dependent transient photocurrent characteristics of the cells with or without AMPA modification.

Another challenge is related to the ZnO-nps itself because ZnO-nps synthesized from aqueous solutions generally has many deep surface defects caused by the residue of hydroxyl groups.[19] It is apparent that these defects are detrimental to the charge transport of the cell.[20] Here, driven by the esterification mechanism, we introduced aminomethyl phosphonic acid (AMPA) molecules onto the ZnO-nps surface to react with these hydroxyl groups. As predicted by DFT calculations (Figure 3a), by releasing a H$_2$O molecule as well as a total energy of about 2.16 eV, the AMPA would bind to the ZnO surface. After this treatment, signals corresponding to the stretching vibration of hydroxyl groups located at about 3300 cm$^{-1}$ were no longer observed in the FT-IR spectra,[37] indicating that the AMPA surface treatment has effectively removed the detrimental hydroxyl groups (Figure 3b). In the meanwhile, a new vibration band centered at about 1076 cm$^{-1}$,[38] corresponding to the stretching vibration of Zn-O-P, appeared in the FTIR spectrum of the ZnO-nps/AMPA sample, indicating the transformation from Zn-O-H to Zn-O-P-N structures, as depicted in Figure 3a. O 1s and Zn 2p XPS spectra also supported the appearance of this new structure (Figure 3c, S9).[39]

However, despite of the surface atomic structure transformation, the distribution of density of states of the ZnO has not exhibited obvious change, and especially the defect states in the forbidden gap were still observed (Figure S10). That is, chemical passivation of the surface defects has not been realized, primarily because the surface Zn was still linked to excessive oxygen atoms. Nonetheless, we found that through the Zn-O-P-N structures, a permanent dipole with a magnitude of 1.16 D was formed perpendicular to the ZnO-np surface, in which negative charge located closer to surface, as shown in Figure 3a. The formation of this dipole was confirmed by surface work function measurement using Kelvin probe force microscopy (KPFM, Figure 4a, b), which showed that the work function of the-ZnO-nps film was decreased by about 280 meV after the AMPA treatment. This permanent dipole would form a local electric field in the surface region, electrostatically repulsing electrons transported in the ZnO-nps away from the defect position, that is forming field-effect passivation toward those point defects.[40,41]

This passivation effect was confirmed using steady-state PL measurement. As in Figure 3d, for the pristine ZnO-nps, significant defect emission at about 620 nm was observed under the ultraviolet excitation, while after the AMPA treatment, the defect emission was reduced by about four times and the emission at about 370 nm corresponding to the ZnO band-edge was significantly enhanced.[26,42] Photocurrent decay (Figure 3e) and EQE (Figure S11) of the solar cells measured under different bias voltages also supported this result. For the pristine ZnO-nps, significant charge loss was observed when the external bias voltage changed from -1 V to 0 V. This arose because -1 V could provide considerable large electric field to extract the photo-generated electrons, reducing its capture by the surface defects, while under 0 V, most of the photo-generated electrons were captured by these surface defects. In contrast, after the AMPA treatment, this charge loss phenomenon was almost completely eliminated, implying that the ZnO-nps induced electron capturing has been effectively passivated.

**Band alignment at the ZnO-nps/AgNWs interface**

Energy band matching is also a critical requirement for the fabrication of high-efficient solar cell. Through band-edge position measurements (Figure 4a-b, Figure S12, S13, S14, Note S2), a mismatch in the conduction band minimum (CBM) was found between ZnO-nps and the Ag NWs. As shown in Figure 4c, CBM of the ZnO-nps was 0.16 eV lower than that of the Ag NWs, which would hinder the fast transfer of electrons across this interface. Interestingly, we found that the dipole introduced by the AMPA has also improved the energy band alignment of this interface. Specifically, the negative charge in the dipole raised the CBM position of the ZnO-nps, thus eliminating the energy barrier at this interface. This improvement helped to obtain an Ohmic contact at the ZnO-nps/Ag NWs interface, as demonstrated by the current-voltage characteristics measurement of a junction comprised of Mo/ZnO-nps/AgNWs/Ag. When fabricating into a complete solar cell, this improved interfacial contact has led to an obvious reduction in the series resistance ($R_S$, shown in Figure 4e). Especially at low temperatures, the $R_S$ was decreased by about four times, and has reached a level comparable to that of the sputtered device. Temperature-dependent $R_S$ measurement further showed that the energy barrier for the interfacial charge transfer has been decreased from 150 meV to 116 meV.[43] Further investigation using modulated transient photocurrent/voltage (m-TPC/TPV) measurements revealed a significant improvement in the internal quantum efficiency (IQE) of the device, showing an enhanced carrier utilization.[34,44] These benefits of the AMPA, including field-effect passivation and energy level alignment improvement, resulted in a considerable reduction in the charge loss in the cell and thus helped us further improved the averaged efficiency of cell from 12.8% to about 14.0%.

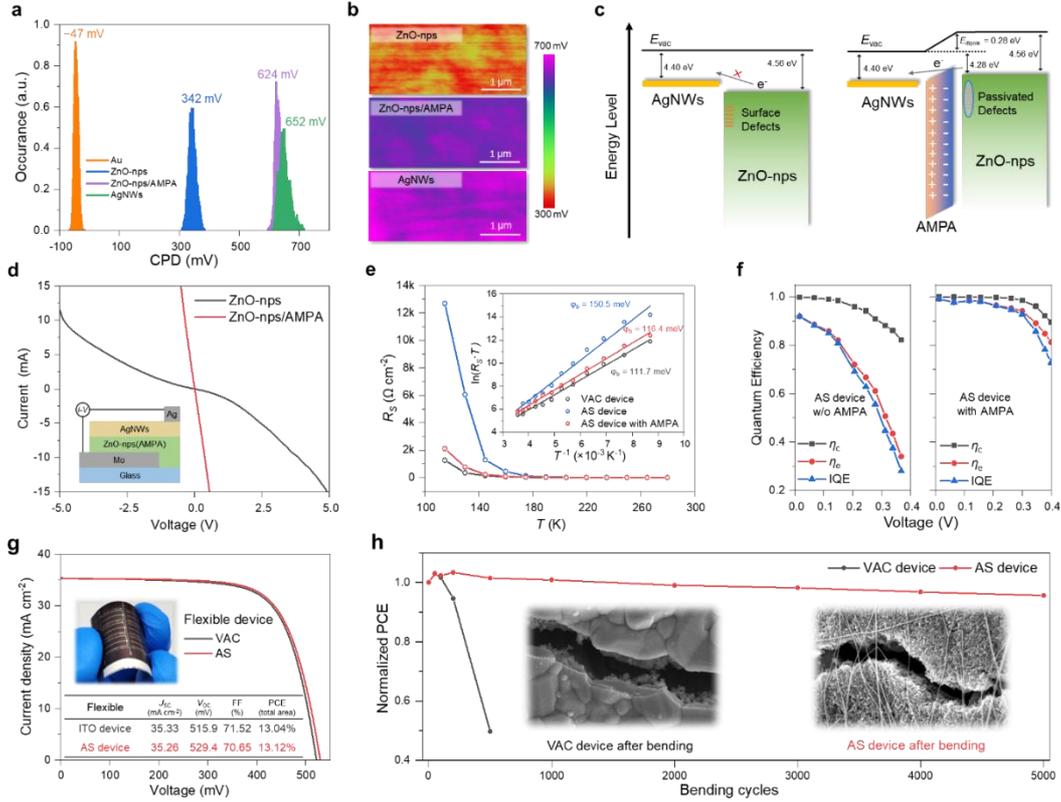

**Figure 4. Band alignment improvement and flexible solar cells.** (a) Contact potential difference (CPD) distribution and (b) mapping images of ZnO-nps, ZnO-nps/AMPA and AgNWs films on ACZTSSe/CdS substrate measured using KPFM. (c) Energy band diagram of ZnO-nps/AgNWs interface with or without the AMPA induced dipole. (d) I-V curve of a junction (inset diagram) comprised of Mo/ZnO-nps/AgNWs/Ag. (e) Temperature-dependent series resistance ($R_S$) of the sputtered cell and the AS cell with or without AMPA modification. The inset figure shows the Arrhenius plots to derive the charge transport barrier $\varphi_b$. (f) The carrier collection efficiency $\eta_C$, extraction efficiency $\eta_e$ and internal quantum efficiency (IQE) of the AS cell with or without the AMPA modification. (g) I-V curve of champion flexible cells fabricated by the conventional sputtering or all-solution method. (h) PCE evolutions of these two types of flexible cells bended for various cycles at a bending radius of 8 mm. The inset SEM figures are sputtered and AS device after being bended for 500 and 5000 cycles, respectively.

**Flexible Kesterite solar cells**

We further used these all-solution technologies to fabricate flexible Kesterite solar cells on molybdenum foil substrate. As shown in Figure 4g, the cell exhibited a high efficiency of 13.12%, which is comparable to that of the sputtered technology and has also reached the highest result reported so far (Table S6).[45,46] In addition to the efficiency, the flexible cell demonstrated excellent bending performance. When bended at different radius ranging from 25 to 5 mm (Figure S15), little efficiency change was observed in the all-solution cell, while obvious efficiency decline has already appeared in the sputtered cell when it was bended at 15 mm. Furthermore, the all-solution cell can sustain 95% of its initial efficiency after being bended for 5000 times at a bending radius of 8 mm (Figure 4h), while the efficiency of the sputtered cell had dropped to 50% of its initial performance after just being bended for 500 times. This significant difference in the flexibility performance primarily lies in the influence of bending on the lateral charge transport. For the sputtered cell, bending caused cracks of the ITO layer, which significantly damaged the lateral charge transport of the cell; while for the all-solution cell, although cracks also appeared after the bending, the long AgNWs can still connect the domains separated by these cracks, thus assuring the lateral charge transport and collection.

These results demonstrate that through the all-solution approach, we can create high-efficiency and robust Kesterite solar cells at extremely low time and energy costs. Due to the similarity in geometric and energy band structure between CZTSSe and Cu(In, Ga)Se (CIGS) solar cells, our strategy reported here should also be able to directly transplant into the preparation of CIGS solar cells, which can effectively reduce equipment investment and production costs in the CIGS industry and additionally help to shorten the production cycle to increase production capacity.

**Conclusion**

In this study, we are committed to developing cost-effective all-solution window layer technology to achieve high performance Kesterite solar cells. Specifically, we have developed a ZnO-nps/AgNWs composite layer that realized comparable light transmission and charge conducting performance to that of the sputtered window layer. To address the issues associated with the film conformal deposition, surface defect and interfacial energy level mismatch of this all-solution window layer, we have introduced multi-interface engineering taking advantages of polymer binders and functional molecules. The polymer, PAA, modified the surface energy of the CdS layer and enhanced its interactions with the ZnO-nps, thereby realizing the conformal deposition of ZnO-nps layer. The functional molecule, AMPA, can react with the hydroxyl group in the ZnO-nps, forming a dipole, which on one hand can field-effect passivate the surface defects of ZnO-nps and on the other hand can improve the energy level alignment at the ZnO-nps/Ag NWs interface. These collective benefits of the multi-interface engineering have helped us to achieve 14.3% and 13.12% efficiency in the all-solution processed rigid and flexible Kesterite solar cells, respectively. Overall, this work not only provides an efficient and low-cost technical route for the future commercialization of CZTSSe solar cells, but also offering important implications for the solution preparation of window layers for other optoelectronic devices.

**Acknowledgements**

This work was supported by the National Natural Science Foundation of China (Grant nos. 52222212 (J. S.), U2002216 (Q. M.), 52172261 (Y. L.), 52227803 (Q. M.), 51972332 (H.W.)). J. S. also gratefully acknowledges the support from the Youth Innovation Promotion Association of the Chinese Academy of Sciences (2022006).